\documentclass[preprint,epsf,aps]{revtex4}
\usepackage{graphicx}
\begin{document}

\title{Normal-state transport properties of PrFeAsO$_{1-x}$F$_y$ superconductor}

\author{D. Bhoi and P. Mandal}
\affiliation{Saha Institute of Nuclear Physics, 1/AF Bidhannagar, Calcutta 700 064, India\\}
\author{ P. Choudhury}
\affiliation{Central Glass and Ceramic Research Institute, 196 Raja
S. C. Mullick Road, Calcutta 700 032, India\\}
\date{\today}
\begin{abstract}

We have synthesized oxygen-deficient fluorine-doped samples of
nominal composition PrFeAsO$_{1-x}$F$_y$   to study the normal and
superconducting state properties.  Resistivity of undoped PrFeAsO
exhibits a strong anomaly at around 155 K due to the
spin-density-wave instability. Fluorine doping ($x$=0.4, $y$=0.12)
suppresses this magnetic instability and drives the system to the
superconducting ground state with  superconducting onset temperature
50 K. The behavior of normal-state resistivity changes from
$T^2$-like to $T$-linear to sublinear in $T$ with increasing
temperature indicating that both electron-electron interaction and
electron-phonon interaction are strong. From the analysis of $T$
dependence of $\rho$, we have estimated  the electron-phonon
coupling strength ($\lambda$) to be quite large (1.3).


\vskip 1cm
\end{abstract}

\maketitle
\newpage

{\bf 1. Introduction}

The recent discovery of superconductivity in iron-based oxypnictide
LaFeAsO$_{1-x}$F$_x$ with transition temperature ($T_c$) 26 K has
generated tremendous activity in this field \cite{kami1,kami2}.
Apart from the high transition temperature, this system exhibits
many interesting properties possibly due to the  presence of iron.
The undoped parent compound LaFeAsO is nonsuperconducting metal and
shows spin-density-wave (SDW) instability below 150 K. Partial
substitution of fluorine for oxygen suppresses the long-range
magnetic ordering and drives the system to superconducting state
above a critical value of fluorine content ($x$). Immediate after
the first report on superconductivity in LaFeAsO$_{1-x}$F$_x$,
attempts have been made to change the composition with an aim to
increase the superconducting transition temperature. Several groups
have shown that the replacement of La by other rare earth elements
($R$) such as Ce, Pr, Nd, Sm, etc increases $T_c$ significantly
\cite{chen1,chen2,ren1,ren2,ren3}. $T_c$ as high as 55
K has been reached in SmFeAsO$_{1-x}$F$_x$ \cite{ren3}.   \\

Usually, two methods have been followed for the preparation of
iron-based oxypnictides. Using high pressure synthesis technique,
one can obtain fluorine-free oxygen-deficient $R$FeAsO$_{1-\delta}$
superconducting samples \cite{ren1,ren4,hiro}. However, samples
prepared in this method contain appreciable amount of impurity
phases resulting from the unreacted ingredients
\cite{ren1,ren4,hiro}. In the other method, oxygen is partially
replaced by fluorine and the sample is prepared either in high
vacuum or in the presence of inert gas. Though samples prepared in
this method are superior in quality as compared to that obtained
from high pressure synthesis, often small amounts of  impurity
phases are present \cite{sefa,ding}. We use a slightly different
method for the preparation of good quality superconducting
PrFeAsO$_{1-x}$F$_y$ samples.  The nominal composition for this
sample is oxygen deficient and oxygen is partially replaced by
fluorine as well ($x \not=y \not= $0). To some extent, this is
basically a combination of above two methods but the samples  can be
prepared in vacuum. In this sample, no impurity has been noticed
from the x-ray diffraction pattern and the superconducting
transition is sharp. We observe that single phase samples
with $x$$\leq$0.4 can be prepared in this method.  \\

In this paper, we report the results on structural, magnetic and
transport properties of PrFeAsO$_{1-x}$F$_y$ as a function of
temperature. In order to extract the normal-state parameters useful
for understanding the mechanism of superconductivity, resistivity
has been measured up to 475 K. The resistivity behavior above $T_c$
shows three distinct temperature regimes wherein two different kinds
of scattering processes originating from the electron-electron
scattering and the electron-phonon scattering are predominant. From
the analysis of high-temperature resistivity, we have estimated the
electron-phonon coupling strength in this system to be large, which
is consistent with the reported large value of normalized energy gap
parameter 2$\Delta/kT_c$=8 \cite{dubro}.\\

{\bf 2. Sample preparation and experimental techniques}

Polycrystalline samples of nominal compositions PrFeAsO and
PrFeAsO$_{0.6}$F$_{0.12}$   were synthesized by conventional solid
state reaction method. High purity chemicals Pr (99.9$\%$), Fe
(99.998$\%$), Fe$_2$O$_3$ (99.99$\%$), AS (99.999$\%$), PrF$_3$
(99.9$\%$) and Pr$_6$O$_{11}$ (99.99$\%$) from Alfa-Aesar were used
for the sample preparation.   Finely grounded powders of
Pr$_{0.96}$As, Fe, Fe$_2$O$_3$, Pr$_6$O$_{11}$ (pre-heated at 600
$^o$C) and PrF$_3$ were thoroughly mixed in appropriate ratios and
then pressed into pellets. The pellets were wrapped with Ta foil and
sealed in an evacuated quartz tube. They were then annealed at 1250
$^o$C for 36 h. Pr$_{0.96}$As was obtained by slowly reacting Pr
chips and As pieces first at 850 $^o$C for 24 h and then at 950
$^o$C for another 24 h in an evacuated quartz tube. The product was
reground, pressed into pellets and then sealed again in a quartz
tube and heated at 1150 $^o$C for about 24 h. The phase purity and
the room-temperature lattice parameters of the samples were
determined by powder x-ray diffraction (XRD) method with Cu
K$_\alpha$ radiation. dc magnetization measurements were done using
a Quantum Design Physical Property Measurement System (PPMS) at a
field of 50 Oe. Resistivity was measured by standard four-probe
technique up to 475 K with an excitation current 5 mA. Electrical
contacts were made using conducting silver paint. Resistivity
measurements above 300 K were done in vacuum in order to reduce the
oxidation of the sample. Thermopower measurements were done using
standard dc differential technique \cite{man}. In this method, a
small temperature gradient was created across the sample and the
corresponding voltage drop was measured.\\

{\bf 3. Experimental results and discussion}

The XRD pattern for the  fluorine-doped PrFeAsO$_{0.6}$F$_{0.12}$
sample is shown in Fig. 1, which could be well indexed on the basis
of tetragonal ZrCuSiAs-type structure with the space group P4/nmm.
We did not observe any impurity phase within the resolution of an
x-ray. The lattice parameters obtained from the Rietveld refinements
are $a$=3.9711 $\AA$ and $c$=8.5815 $\AA$. Fe-Fe bond length
calculated from the refinements is 2.8080 $\AA$. We have also
determined the lattice parameters for PrFeAsO. For this sample, both
$a$ (=3.984 $\AA$) and $c$ (=8.596 $\AA$) are larger than that for
the fluorine-doped sample.  As expected, the incorporation of F in
place of O reduces the size of the unit cell. The above values of
lattice parameters are comparable with those reported for 12$\%$
F-doped and PrFeAsO samples, respectively \cite{ren1}. We would like
to mention here that the nominal oxygen-site vacancy in our present
sample is $(1-x-y)$=0.28. However, we are unable to determine the
exact numbers of the oxygen-site vacancies from x-ray refinements
because x-ray is insufficient for the accurate determination of the
light atom stoichiometry.  Neutron diffraction may reveal the
information on oxygen-site vacancy. Nevertheless, one can prepare
good quality and single phase samples using this technique. \\

To test the bulk nature of superconductivity, we have measured the
temperature dependence of zero-field cooled (ZFC) and field cooled
(FC) magnetic susceptibility for the F-doped sample at 50 Oe (Fig.
2). Both ZFC and FC susceptibilities start to deviate from the
normal behavior and become negative below 48 K and, their values
increase with decreasing $T$.  The shielding and Meissner fractions
were calculated from the ZFC and FC data, respectively. Assuming
theoretical density  6.96 g/cm$^3$, a shielding fraction of 65$\%$
and a Meissner fraction of 23$\%$ are found at 5 K. These volume
fractions are comparable with those (74$\%$ and 33$\%$,
respectively) observed in LaFeAsO$_{0.89}$F$_{0.11}$ at 2 K for
$H$=20 Oe \cite {sefa}. The present values for volume fractions
might enhance significantly, if the measurements could have been
performed at lower fields. \\

Figure 3(a) shows the temperature dependence of resistivity for
PrFeAsO$_{1-x}$F$_y$ samples.  The behavior of $\rho$($T$) for the
parent sample is qualitatively similar to that of LaFeAsO
\cite{kami1}. An anomalous peak associated with SDW shows up at
$T_s$=155 K. Below the occurrence temperature of SDW,  resistivity
drops steeply.  Above 250 K, $\rho$ increases linearly with $T$. The
linear behavior is maintained up to as high as 475 K. However, at
low temperature $\rho$ exhibits a power-law behavior. For $T<T_s$,
resistivity data can be fitted well with an expression,
$\rho$=$\rho_0 +aT^n$, with $n$ close to 1.5. Fluorine doping leads
to the suppression of the SDW state. For $y$=0.12, the behavior of
$\rho$ vs $T$ curve changes completely and no anomaly due to SDW
formation is observed. $\rho$ for this sample decreases
monotonically with decreasing temperature until the superconducting
onset temperature ($T_{c}^{on}$) is reached below which $\rho$ drops
sharply and becomes zero just below 47 K. The inset displays the
enlarged view of the onset of superconductivity. For this sample,
$T_{c}^{on}$ is 50 K (using the 90$\%$ criterion), and a transition
width ${\Delta}$${T_c}$=$T_c$(90$\%$) $-$ $T_c$(10$\%$)=2.7 K. For
LaFeAsO$_{0.89}$F$_{0.11}$ the value of ${\Delta}$${T_c}$ is 4.5 K
\cite{sefa}. Thus the superconducting transition for
PrFeAsO$_{0.6}$F$_{0.12}$
is sharper than that for LaFeAsO$_{1-x}$F$_x$ \cite{kami1,kami2,sefa}. \\

From the behavior of the $T$ dependence of $\rho$, one can identify
three distinct temperature regimes. We find that $\rho$ above $T_c$
exhibits a quadratic temperature dependence, $\rho$=$\rho_0$+$AT^2$,
in the range of $T$ between 70 K and 170 K [Fig. 3(b)]. The $T^2$
behavior of $\rho$ below 170 K indicates a strong electronic
correlation and is consistent with the formation of a Fermi-liquid
state. The value of $A$ determined from the fit is 2.45
$\times$10$^{-5}$ m$\Omega$ cm K$^{-2}$. This value of $A$ is
comparable with that observed in some semiheavy-fermion compounds
\cite{kado} but larger than that reported for several metallic
oxides \cite{imada}. Often, strongly correlated systems are
characterized by the Kadowaki-Woods (KW) ratio $A$/$\gamma^2$, where
$\gamma$ is the coefficient of linear heat capacity \cite{kado}.
Using the reported value of $\gamma$=4.1 mJ K$^{-2}$ mol$^{-1}$ for
LaFeAsO$_{0.89}$F$_{0.11}$ system \cite{sefa}, we find that the KW
ratio for the present sample is 1.46 $\times$ 10$^{-3}$ $\mu$
$\Omega$ cm K$^{2}$ mol$^{2}$ mJ$^{-2}$. This value of KW ratio is
about two orders of magnitude larger than that for the heavy-Fermion
system (1.0 $\times$ 10$^{-5}$ $\mu$ $\Omega$ cm K$^{2}$ mol$^{2}$
mJ$^{-2}$) but comparable to  that for Na$_{0.7}$CoO$_2$
\cite{kado,li}. Such a large value KW ratio means that the strong
electron correlations are responsible for enhancing not so much the
effective mass of quasiparticles as their scattering rate. A
systematic deviation from $T^2$ to $T$-linear behavior of $\rho$ is
observed above 170 K. Figure 3(c) clearly shows that $\rho$ is
linear over a wide range of $T$ above 170 K. The temperature
coefficient of $\rho$ in the linear regime is 8.6 $\mu$$\Omega$ cm
K$^{-1}$. This value of d$\rho$/d$T$ is an order of magnitude larger
than that observed in high temperature superconductors where $\rho$
exhibits linear behavior up to as high as 1100 K \cite{gur}. With
increasing $T$ above 280 K, $\rho$ increases slower than $T$-linear
and exhibits a downward curvature. Thus the nature of temperature
dependence of resistivity  for the fluorine-doped superconducting
sample is very different from that of non-superconducting parent
compound at high temperatures. Undoped PrFeAsO does not show any
saturation-like behavior at high temperatures. The high temperature
resistivity behavior of the supeconducting sample is quite similar
to that observed in several A15 compounds and Chevrel phases
\cite{marc}. Qualitatively, this reveals that the mean free path $l$
of the present system is comparable with that of interatomic spacing
($a_{in}$) above 300 K. In metals with $l$$\sim$$a_{in}$, $\rho$
displays a strong deviation from linearity at high temperature,
known as resistivity saturation. The linear behavior of $\rho$
between 175 K and 280 K and its deviation from linearity above room
temperature, are the signatures of strong electron-phonon interaction.\\

The temperature dependence of thermopower $S$ for both the samples
are shown in Fig. 4. In the measured temperature range, $S$ is
negative for both the samples. The negative value of $S$ indicates
electron carriers. The magnitude of $S$ for the superconducting
sample is about 3 times larger than that for the nonsuperconducting
sample. For both the samples, $S$ increases approximately linearly
with $T$ for 175 K$\leq$$T$$\leq$ 280 K and a systematic deviation
occurs at high temperature above 280 K. This behavior is somewhat
similar to the $T$ dependence of $\rho$ above 175 K [Fig. 3(c)].
Below 175 K, $S$ decreases with $T$ faster than linear and passes
through a minimum at $T_{min}$. For the nonsuperconducting sample,
$S$ increases at a much faster rate than that for the
superconducting sample below $T_{min}$ and exhibits a sharp maximum
at 120 K. This rapid increase of $S$ below $T_{min}$ is possibly due
to the formation of SDW state. We observe that below 175 K, $S$ for
the superconducting sample can be fitted well with an expression $S
\sim (T-T_{min})^2$. For LaFeAsO$_{0.89}$F$_{0.11}$, $S$ is large
and negative, and the nature of $S$($T$) curve is qualitatively
similar to that of PrFeAsO$_{0.6}$F$_{0.12}$ sample \cite{sefa}.
Indeed, $S$ for LaFeAsO$_{0.89}$F$_{0.11}$ is slightly larger than
that for the present sample and exhibits an approximate linear
behavior below 35 K, as expected from Fermi liquid theory. The
smaller value of $S$ for PrFeAsO$_{0.6}$F$_{0.12}$ may be an
indication of higher Fermi energy for this sample. Due to the higher
superconducting transition temperature, no linear region is,
however, expected to appear for PrFeAsO$_{0.6}$F$_{0.12}$ sample.\\

We now combine resistivity results with band-theory parameters in
order to estimate the strength of electron-phonon coupling
($\lambda$). Expressing the resistivity in terms of plasma energy
($\hbar\omega_p$) and electron-phonon scattering time ($\tau_{ep}$),
$\rho(T)-\rho(0)$=
4$\pi/(\omega_p^2\tau_{ep})=4\pi{v_F}/(\omega_p^2{l})$, where $v_F$
is the Fermi velocity \cite{gur,allen}. Also, at high temperature
where electron-phonon scattering dominates resistivity, $\tau_{ep}$
is given by $\hbar/\tau_{ep} =2\pi\lambda{kT}$ \cite{allen}. From
these relations we can deduce

\begin{equation}
\lambda={\frac{\hbar\omega_p^2}{8\pi^2k}}{\frac{d\rho}{dT}}
\end{equation}

and

\begin{equation}
\lambda={\frac{{\hbar}v_F}{2\pi l k T}}.
\end{equation}

Assuming $l$$\sim$ $a$=4 $\AA$ at 300 K and the value of Fermi
velocity $v_F$=1.3$\times$$10^7$ cm/s calculated from band theory
\cite{sing}, we find a large value for $\lambda$=1.3. $\lambda$ can
also be estimated independently using the values of resistivity
slope (8.6 $\mu\Omega$ cm/K) in the linear region and plasma energy
($\hbar\omega_p$$\sim$ 0.8 eV) determined from the inplane
penetration depth \cite{drew}.  In this case also $\lambda$ is 1.3.
Thus the value of $\lambda$ calculated in two different methods is
same and quite large. This suggests that electron-phonon scattering
is dominating the high temperature resistivity of the
superconducting sample and is consistent with the resistivity
saturation.\\

{\bf 4. Conclusions}

 In conclusion, we have analyzed resistivity and
thermopower for oxygen-deficient fluorine-doped samples of nominal
composition PrFeAsO$_{1-x}$F$_y$ over a wide range of temperature.
Fluorine doping ($x$=0.4, $y$=0.12) suppresses the formation of SDW
state and drives the system to the superconducting ground state with
$T_c$=50 K. With increasing $T$, resistivity above $T_c$ crosses
over from $T^2$ dependence due to the electron-electron interaction
to linear in $T$ and then to a saturation-like behavior at higher
temperature due to the electron-phonon interaction. In contrast,
undoped PrFeAsO exhibits linear $T$ dependence up to 475 K. The
occurrence of saturation-like behavior indicates that the
electron-phonon interaction is strong.  We have estimated the
electron-phonon coupling parameter $\lambda$$\sim$ 1.3 from the
linear and sublinear dependence of $\rho$ for the superconducting
sample. This result together with reported large value of normalized
energy gap parameter 2$\Delta/kT_c$=8 suggests that this system
belongs to the class of strong coupling superconductors.\\

{\bf Acknowledgement}

The authors would like to thank B. Ghosh for stimulating discussions
and S. K. De, M. Patra, J. Ghosh and A. Pal for the technical help
during the sample preparation and
measurements.\\

\newpage

\begin{figure}[htb]
\begin{center}
\includegraphics[scale=2.3,angle=0]{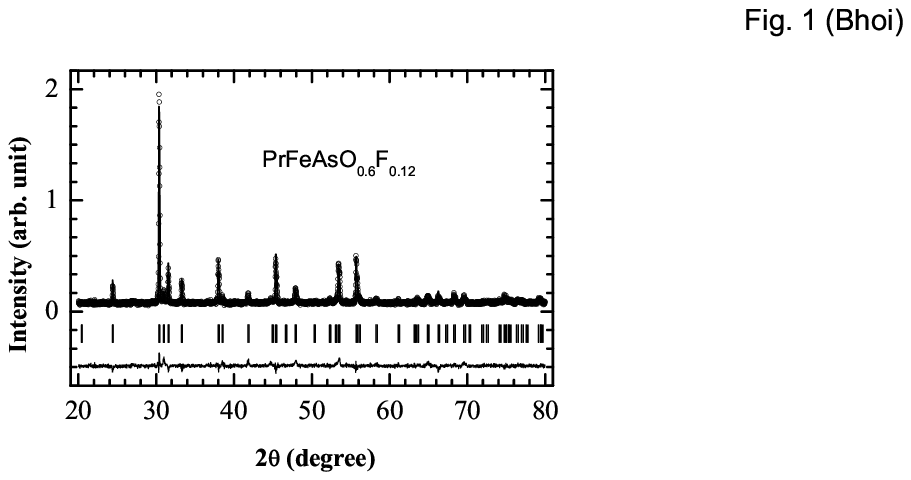}
\caption{The x-ray diffraction pattern for the superconducting
PrFeAsO$_{0.6}$F$_{0.12}$  sample. The solid line corresponds to
Rietveld refinement of the diffraction pattern with $P4/nmm$ space
group.}
\label{bhoi1}
\end{center}
\end{figure}

\begin{figure}[htb]
\begin{center}
\includegraphics[scale=2.4,angle=0]{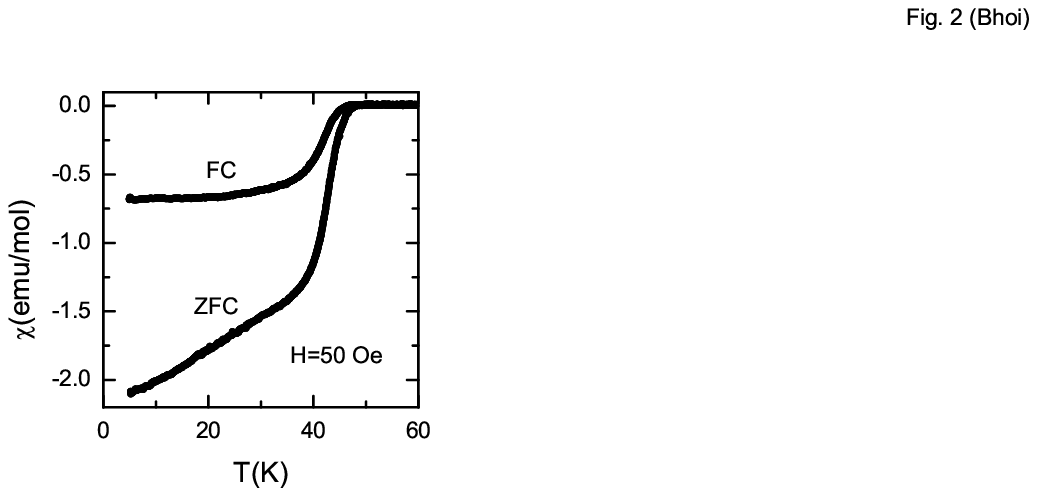}
\caption{ Temperature dependence of FC  and ZFC  dc magnetic
susceptibility ($\chi$)  measured at a field of $H$=50 Oe for
PrFeAsO$_{0.6}$F$_{0.12}$ sample.}
\label{bhoi2}
\end{center}
\end{figure}

\begin{figure}[htb]
\begin{center}
\includegraphics[scale=1.3,angle=0]{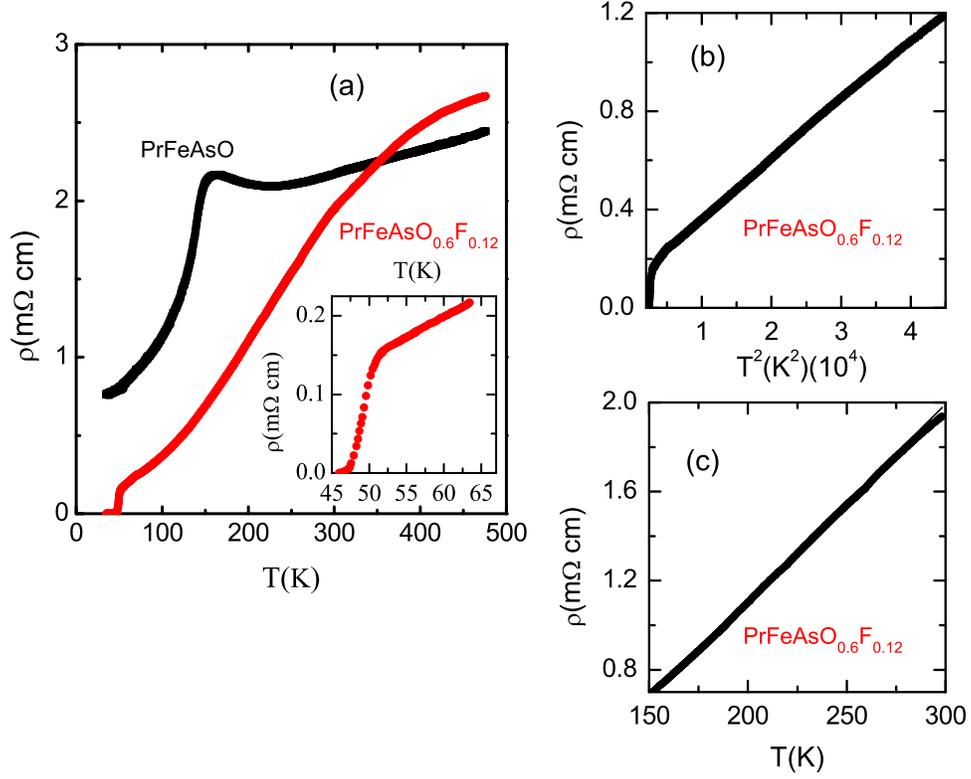}
\caption{ (a)Temperature dependence of resistivity ($\rho$) for
PrFeAsO$_{0.6}$F$_{0.12}$ and PrFeAsO samples.Inset(a): Enlarged
view of resistivity change close to the superconducting transition
temperature.(b)$T^2$ dependence of $\rho$ for superconducting
PrFeAsO$_{0.6}$F$_{0.12}$ sample in the range 70 K$\leq$ 170
K.(c)Linear behavior of $\rho$ for 175 K$\leq T \leq$280 K for
PrFeAsO$_{0.6}$F$_{0.12}$ sample}
\label{bhoi}
\end{center}
\end{figure}


\begin{figure}[htb]
\begin{center}
\includegraphics[scale=2.4,angle=0]{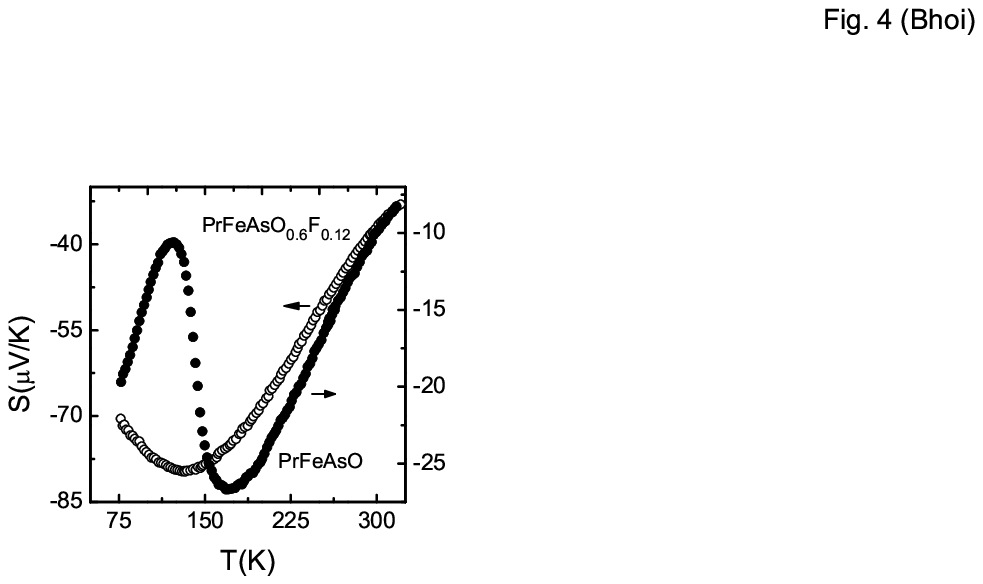}
\caption{ Temperature dependence of thermopower ($S$) for
PrFeAsO$_{0.6}$F$_{0.12}$ (open symbol) and PrFeAsO (solid symbol)
samples.} \label{bhoi4}
\end{center}
\end{figure}

\end{document}